\documentstyle{article}

\newcommand{\be}{\begin{equation}}
\newcommand{\ee}{\end{equation}}
\newcommand{\order}{{\cal O}}

\newcommand{\nr}{{\rm NR}}
\newcommand{\pv}{{\bf p}}

\newcommand{\fig}[1]{Fig.~\ref{#1}}
\newcommand{\Ref}[1]{Ref.~\cite{#1}}
\newcommand{\kin}{{\rm kin}}
\newcommand{\eq}[1]{Eq.~(\ref{#1})}

\newcommand{\Mbz}{{M_b^0}}

\newcommand{\msbar}{{\rm \overline{MS}}}
\newcommand{\NRQCDcoll}{
C.~T.~H.~Davies,$^a$ K.~Hornbostel,$^b$ A.~Langnau,$^c$
G.~P.~Lepage,$^c$\\ A.~Lidsey,$^a$ C.J.~Morningstar,$^d$
J.~Shigemitsu,$^e$ J.~Sloan,$^f$\\[.4cm]

\small $^a$University of Glasgow, Glasgow, UK G12 8QQ \\
\small $^b$Southern Methodist University, Dallas, TX 75275 \\
\small $^c$Newman Laboratory of Nuclear Studies, Cornell University,
Ithaca, NY 14853 \\
\small $^d$University of Edinburgh, Edinburgh, UK EH9 3JZ \\
\small $^e$The Ohio State University, Columbus, OH 43210 \\
\small $^f$Florida State University, SCRI, Tallahassee, FL 32306 \\
}

\begin{document}

\title{ A New Determination of $M_b$ Using Lattice QCD}

\author{
\NRQCDcoll \\ }

\date{\small Final Revision\,---\,September 1994}

\maketitle

\begin{abstract}
Recent results from lattice QCD simulations provide a realistic picture,
based upon first principles, of~$\Upsilon$ physics. We combine these results
with the experimentally measured mass of the $\Upsilon$~meson  to obtain an
accurate and reliable value for the $b$-quark's pole mass. We use two
different methods, each of which yields a mass consistent with
$M_b = 5.0(2)$~GeV.  This corresponds to a bare mass of
$M_b^0 = 4.0(1)$~GeV in our lattice
theory and an $\msbar$~mass of $M_b^\msbar(M_b)=4.0(1)$~GeV.
We discuss the implications of this result for the $c$-quark
mass. 
\end{abstract}

\vspace{.2cm}

The mass~$M_b$ of the $b$~quark is a fundamental parameter of the Standard
Model.
Various phenomenological studies
have suggested masses in the range of 4.5--5~GeV \cite{mb-refs}.
In this paper we
use  new results from lattice QCD to obtain a value for $M_b$
from the measured mass of the $\Upsilon$~meson \cite{nrqcd-results}. We
employ two methods for computing the quark mass; both are consistent with
$M_b=5.0(2)$~GeV, which is presently among the most accurate and reliable
of determinations.

The notion of quark mass is complicated by quark confinement~\cite{pbm-chat}.
In a particular process, a $b$-quark's effective, or running, mass~$M_b(q)$
depends upon the typical momentum~$q$ transferred to it. In
perturbation theory, at least, this mass stops running when $q$ falls
below~$M_b(q)$. Below this point, $M_b(q)$ equals the quark's pole mass,
which is defined perturbatively, in terms of the bare mass, by the pole
in the quark propagator. It is this mass we quote for $M_b$ above. The
use of perturbation theory is certainly justified when $q\!=\!M_b$. At
very low $q$'s the effective mass may be modified away from our value
by nonperturbative effects, but
our value should hold for $\Lambda_{QCD}\!\ll\!q\!\ll\!M_b$. It is
therefore the appropriate mass to use in quark potential models and other
phenomenological applications that involve this momentum range.
We may also use perturbation theory to relate our bare quark mass to the
$\msbar$~mass; we find $M_b^\msbar(M_b) = 4.0(1)$~GeV.

Both of our mass determinations rely on numerical simulations of the
$\Upsilon$ using lattice QCD. In lattice QCD spacetime is approximated by
a discrete grid of finite physical
volume. An action for QCD is defined in terms of quark and gluon
fields on the nodes of the grid and the links joining them.
Monte Carlo methods are used to evaluate the path integrals that define
various  correlation, or Green's, functions. The energies and wavefunctions
of the~$\Upsilon$ and its excitations are extracted from these correlation
functions.

Our gluon configurations were generated using the standard Wilson
lagrangian for gluons\cite{std-ref}.
For most of our simulations we used a $16^3\times24$~grid with
$\beta\equiv 6/g^2=6$~\cite{kilcup}, where $g$ is the bare coupling
constant. This corresponds to an inverse lattice spacing of
$a^{-1} = 2.4(1)$~GeV and a volume of
approximately $1.3^3\times2.0~{\rm fm}^4$.
The lattice volume is more than adequate since $\Upsilon$'s have a radius of
only $R_\Upsilon \approx 0.2$~fm. The finite grid spacing introduces
systematic errors of order $a^2/R_\Upsilon^2$; we have estimated
these to be less than~5\% (for binding energies) both from quark potential
models  and from other simulations with smaller lattice
spacings~\cite{ukqcd}.

The gluonic configurations described above do not include effects due to
light-quark vacuum polarization. There are strong theoretical arguments
suggesting that $\Upsilon$~physics is largely unaffected by the light
quarks, at least for states that are well below the $B\overline
B$~threshold. To verify this we repeated a part of our analysis for gluon
configurations  that contain contributions from $n_f\!=\!2$~flavors of
light quark~\cite{urs}. The mass of the light quarks in these
configurations (under 100~MeV) is larger than the mass of a
$u$~or $d$~quark but negligible relative to the typical momentum transfers in
an~$\Upsilon$ (about 1~GeV). Thus the quarks are effectively massless.
The lattice spacing and volume are the same for these
configurations as for the $n_f\!=\!0$ configurations described above.
The bulk of the results presented in this paper are based upon the
simulations with~$n_f\!=\!0$; however, we found these essentially
unchanged by the inclusion of light-quarks~\cite{unquenched}.

We used the NRQCD lagrangian for the $b$ quarks \cite{nrqcd-ref}. This is a
nonrelativistic lagrangian that is tuned to efficiently reproduce results
from  continuum relativistic QCD order-by-order in the lattice
spacing~$a$ and quark velocity~$v$. We included the leading
relativistic and finite-lattice-spacing corrections; further
corrections are almost certainly negligible for our purposes.
The only parameter in this
lagrangian, other than the QCD charge $g$, is the bare quark
mass~$M^0_b$. This is tuned so that the simulation gives the correct
dispersion relation for the~$\Upsilon$. The energy of a
low-momentum~$\Upsilon$ can be computed in the simulation, and has the form
 \be \label{disp}
 E_\Upsilon(\pv)\approx E_\nr(\Upsilon)\: +\: \pv^2/\;
                                               2 M_\kin(\Upsilon) \, ,
 \ee
where $E_\nr(\Upsilon)$ is related to the nonrelativistic binding energy of
the meson, and $M_\kin(\Upsilon)$ is the kinetic mass of
the meson. We tuned the bare quark mass~$\Mbz$ until
$M_\kin(\Upsilon)$ was equal to the measured mass of
the $\Upsilon$.\footnote{In a purely
nonrelativistic theory, Galilean invariance implies that
the kinetic mass of a bound state equals the sum of
the masses of its constituents. In a relativistic theory the binding
energy enters as well, resulting in a 10\% shift for $\Upsilon$'s.
Thus it is important to include relativistic corrections in the quark
action when tuning the quark mass using~$M_\kin$.}
We found that the correct bare mass is given by~$a\,M_b^0=1.7(1)$.

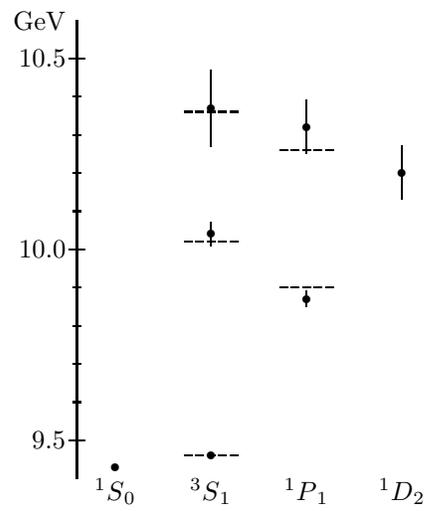
\begin{figure}[p]
\begin{center}
\setlength{\unitlength}{.02in}
\begin{picture}(80,140)(0,930)
\put(15,940){\line(0,1){120}}
\multiput(13,950)(0,50){3}{\line(1,0){4}}
\multiput(14,950)(0,10){10}{\line(1,0){2}}
\put(12,950){\makebox(0,0)[r]{9.5}}
\put(12,1000){\makebox(0,0)[r]{10.0}}
\put(12,1050){\makebox(0,0)[r]{10.5}}
\put(12,1060){\makebox(0,0)[r]{GeV}}
\put(25,940){\makebox(0,0)[t]{${^1S}_0$}}
\put(25,943){\circle*{2}}

\put(50,940){\makebox(0,0)[t]{${^3S}_1$}}
\multiput(43,946)(3,0){5}{\line(1,0){2}}
\put(50,946){\circle*{2}}

\multiput(43,1002)(3,0){5}{\line(1,0){2}}
\put(50,1004){\circle*{2}}
\put(50,1004){\line(0,1){3}}
\put(50,1004){\line(0,-1){3}}

\multiput(43,1036)(3,0){5}{\line(1,0){2}}
\put(50,1037){\circle*{2}}
\put(50,1037){\line(0,1){10}}
\put(50,1037){\line(0,-1){10}}

\put(75,940){\makebox(0,0)[t]{${^1P}_1$}}

\multiput(68,990)(3,0){5}{\line(1,0){2}}
\put(75,987){\circle*{2}}
\put(75,987){\line(0,1){2}}
\put(75,987){\line(0,-1){2}}

\multiput(68,1026)(3,0){5}{\line(1,0){2}}
\put(75,1032){\circle*{2}}
\put(75,1032){\line(0,1){7}}
\put(75,1032){\line(0,-1){7}}

\put(100,940){\makebox(0,0)[t]{${^1D}_2$}}
\put(100,1020){\circle*{2}}
\put(100,1020){\line(0,1){7}}
\put(100,1020){\line(0,-1){7}}

\end{picture}
\end{center}
 \caption{NRQCD simulation results for the spectrum of the
$\Upsilon$ system including radial excitations.
  Experimental
 values (dashed lines) are indicated for the triplet
$S$-states, and for the
 spin-average of the triplet $P$-states. The energy zero from
 simulation results is adjusted to give the correct mass to the
 $\Upsilon(1{^3S}_1)$. These data are for $n_f\!=\!0$.}
\label{spect}
\end{figure}

\begin{figure}[p]
\begin{center}
\setlength{\unitlength}{0.02in}
\begin{picture}(110,140)(0,-50)
\put(15,-50){\line(0,1){80}}
\multiput(13,-40)(0,20){4}{\line(1,0){4}}
\multiput(14,-40)(0,10){7}{\line(1,0){2}}
\put(12,-40){\makebox(0,0)[r]{$-40$}}
\put(12,-20){\makebox(0,0)[r]{$-20$}}
\put(12,0){\makebox(0,0)[r]{$0$}}
\put(12,20){\makebox(0,0)[r]{$20$}}
\put(12,30){\makebox(0,0)[r]{MeV}}

\multiput(43,-40)(3,0){5}{\line(1,0){2}}
\put(58,-40){\makebox(0,0)[l]{$\chi_{b0}$}}
\put(50,-34){\circle*{2}}
\put(50,-34){\line(0,1){7}}
\put(50,-34){\line(0,-1){7}}

\put(36,-5){\makebox(0,0)[l]{$h_b$}}
\put(30,-5){\circle*{2}}
\put(30,-5){\line(0,1){1}}
\put(30,-5){\line(0,-1){1}}

\multiput(63,-8)(3,0){5}{\line(1,0){2}}
\put(78,-8){\makebox(0,0)[l]{$\chi_{b1}$}}
\put(70,-10){\circle*{2}}
\put(70,-10){\line(0,1){4}}
\put(70,-10){\line(0,-1){4}}

\multiput(83,13)(3,0){5}{\line(1,0){2}}
\put(98,13){\makebox(0,0)[l]{$\chi_{b2}$}}
\put(90,13){\circle*{2}}
\put(90,13){\line(0,1){3}}
\put(90,13){\line(0,-1){3}}
\end{picture}
\end{center}
 \caption{Simulation results for the spin splittings between the lowest lying
 $P$-wave states in the $\Upsilon$~family.
 The dashed lines are the experimental
 values for the triplet states.  Energies
  are measured relative to the center of  mass of the triplet states.
 These data are for $n_f\!=\!0$. }
\label{fs}
\end{figure}
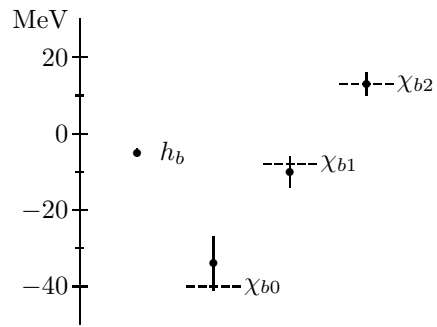

Our simulations very successfully reproduce the general features of
$\Upsilon$ physics. This is illustrated by \fig{spect}, which compares
simulation results for the spectrum of the lowest-lying states
with experiment. In \fig{fs},
simulation results for the fine structure of the lowest lying $P$~state are
compared with experiment. The excellent agreement supports the reliability
of these simulations. We emphasize that these are calculations from
first principles; our approximations can be systematically improved.
The only inputs are the lagrangians describing gluons and quarks, and the
only parameters are the bare coupling constant and quark mass. In particular,
these simulations are {\em not\/} based on a phenomenological quark
potential model.

Our first procedure for computing the $b$-quark mass uses simulation results
for the nonrelativistic energy~$E_\nr(\Upsilon)$ of the $\Upsilon$
(\eq{disp}). The quark mass is
 \be \label{fmeth}
 M_b = (1/2) \, \left(M_\Upsilon - (E_\nr - 2E_0)\right),
 \ee
where $M_\Upsilon = 9.46$~GeV is the experimentally measured mass of
the~$\Upsilon$, and $E_0$ is the nonrelativistic energy of a $b$~quark
with $\pv\!=\!0$ in NRQCD. The quantity $E_\nr(\Upsilon) - 2E_0$ is
the effective binding energy of the meson. The quark energy~$E_0$ is
ultraviolet divergent as the lattice spacing~$a$ vanishes, and so
at small $a$, it depends primarily on physics at
momenta of order~$1/a$. Thus it can be computed using weak-coupling
perturbation theory \cite{colin}, with
 \be
 E_0 = (b_0/a)\: \alpha_V(q_0)\,\left(1 + \order(\alpha_V)\right) \, .
 \ee
Here  $\alpha_V(q_0)$ is the strong coupling constant
defined in \cite{lm}.\footnote{All perturbative results
in this paper are obtained using the techniques and conventions of \cite{lm};
in particular, the definition of the running coupling constant and the
procedure for setting the coupling scale are from this reference. The value
of $\alpha_V$ is determined from the measured expectation value of
the plaquette operator, as discussed in \cite{lm}. This coupling
constant is related to the $\msbar$~coupling by: $\alpha_\msbar(Q)
= \alpha_V(\exp(5/6)\,Q)\left(1+2\,\alpha_V/\pi\right)$.}

In Table~\ref{first-method} we present our simulation results for $E_\nr$,
and the corresponding $E_0$'s from perturbation  theory. We list results
for several values of the bare quark mass; for the $b$-quark,
$aM_b^0 = 1.7(1)$. From these
results we conclude that the pole mass of the $b$ quark is $M_b =
5.0(2)$~GeV. The major sources of uncertainty in this determination are:

 (1) {\it Tuning errors in the bare quark mass.} The binding energy,
$E_{\rm NR} - 2E_0$, like the radial and orbital level
splittings, should be almost independent of the quark's mass for
masses of order $M_b$. Thus $M_b$, as determined from~\eq{fmeth},
should also be independent of the quark's mass.  This is
confirmed by our results in Table~\ref{first-method}.
Errors in $M_b$ due to uncertainty in the bare quark mass are less than 1\%.

 (2) {\it Two-loop corrections to $E_0$.} Quantitative studies \cite{lm} of
the reliability of perturbation theory at scales of order~$q_0$
suggest that two-loop corrections to~$E_0$ could range from
1--30\% of $E_0$. A 20\% contribution would shift~$M_b$ by about 3\%.
Two-loop corrections could not be much larger than this without
ruining the agreement we find between results obtained with widely different
quark masses. Nonperturbative effects are most likely smaller.

 (3) {\it Simulation errors in $E_{\rm NR}(\Upsilon)$.} The dominant
sources of error in $E_{\rm NR}$ are the $\order(a^2)$ errors in the gluon
action, and our use of zero and two flavors of light quarks rather than
three.  Based upon the results in the Table~\ref{first-method},
together with results from simulations with smaller lattice
spacings~\cite{ukqcd}, we expect these effects to shift the quark mass by
less than 100~MeV.  Statistical errors are negligible here.

 (4) {\it Simulation errors in $a^{-1}$.} The lattice spacing is determined
by matching simulation results for the $\Upsilon$~spectrum to
experiment. It is easily measured to within $\pm 5\%$. Since $E_{\rm NR}$
and $2E_0$ are each only about 10\% of $M_\Upsilon$, this amounts to an
uncertainty of less than 0.5\% in $M_b$.
 \begin{table}
 \begin{center}
 \begin{tabular}{lccccccc|c}
  & $\beta$ & $n_f$ & $aM_b^0$ & $aE_\nr(\Upsilon)$ & $b_0$ & $aq_0$
  & $aE_0$ & $M_b$\\
 \cline{2-9}
 full NRQCD & 6.0  &0& 1.71 & 0.453(1) &0.75 &0.48 &0.32 & 4.95 \\
            & && 1.80 & 0.451(1) &0.77 &0.51 &0.31 & 4.92 \\
            & && 2.00 & 0.444(1) &0.81 &0.55 &0.30 & 4.91 \\
            & && 3.00 & 0.412(1) &0.91 &0.68 &0.28 & 4.92 \\
            & 5.6 &2& 1.80 & 0.493(1) & 0.77 & 0.51 & 0.38 & 5.06 \\
 $\delta H=0$ & 6.0&0  & 1.80  & 0.448(1) & 0.76 & 0.47 & 0.33 & 4.98
 \end{tabular}
 \end{center}
  \caption{Simulation and perturbative results used in the first method for
  determining~$M_b$. Values for $M_b$ are in GeV, and are determined
  using~Eq.~(6). Results are presented for $n_f=0$ and 2 light-quark
  flavors, and
  for a range of bare quark masses; the $b$-quark has
  $aM_b^0 = 1.7(1)$.
  Results are
  also given for just the leading term in
  the NRQCD lagrangian ($\delta H = 0$). }
 \label{first-method}
 \end{table}

The relativistic and finite-lattice-spacing corrections
in the quark lagrangian should have only a
small effect on the total mass of the~$\Upsilon$. This is confirmed by our
simulation, as illustrated in Table~\ref{first-method}. The last entry in
the table was obtained from a simulation without these corrections.
It agrees well with results from the full simulation, confirming that
further corrections are probably unimportant.

Our second procedure for determining~$M_b$ is to tune the bare quark
mass~$M_b^0$ until the kinetic mass of the~$\Upsilon$, as computed in the
simulation (\eq{disp}), agrees with the $\Upsilon$'s true mass. Then
the pole mass is $Z_m\,M_b^0$, where the renormalization
constant~$Z_m$ is computed using perturbation theory \cite{colin}, with
 \be
 Z_m = 1 + b_m \alpha_V(q_m) + \order(\alpha_V^2).
 \ee
To reduce the sensitivity of the result to the
values of the lattice spacing and bare quark mass, we rewrite the
expression for the pole mass as
 \be \label{smeth}
 M_b = Z_m\,M_\Upsilon\,\frac{aM_b^0}{aM_\kin(\Upsilon)} \, .
 \ee
Here $M_\Upsilon = 9.46$~GeV is the~$\Upsilon$'s experimental mass, while
$M_\kin(\Upsilon)$ is its mass as determined from a simulation using the bare
mass~$M_b^0$. Our results are summarized in Table~\ref{second-method}.
For $n_f\!=\!0$, the
kinetic mass is closest to the $\Upsilon$~mass when~$aM_b^0 = 1.71$, and
therefore the renormalized quark mass is $M_b = 4.9(2)$~GeV.  The
corresponding bare quark mass is $M_b^0 = 4.1(1)$~GeV (from
\eq{smeth} with $Z_m\!\to\!1$). The $n_f\!=\!2$ data must be
extrapolated slightly in $a M_b^0$
to obtain the correct $\Upsilon$~mass. Using
the $n_f\!=\!0$ data as a guide for this extrapolation, we
obtain $M_b=5.0(2)$~GeV and $M_b^0=4.0(1)$~GeV for $n_f\!=\!2$. The main
sources of uncertainty in these determinations are:

(1) {\it Two-loop corrections to $Z_m$.} The one-loop corrections to
$Z_m$ shift~$M_b$ by 15--20\%. If two-loop corrections are 15--20\% of
the one-loop corrections, they could shift $M_b$ by 2--4\%. The bare
mass is unaffected by these corrections.

(2) {\it Simulation errors in $a M_{\rm kin}(\Upsilon)$.} These are mainly
due to the omission of $\order(v^4)$ relativistic corrections in the quark
action.  Such effects should shift $M_b$ by less than 1\%.
(The $\order(v^2)$~corrections,
which we include, shift~$M_b$ by only 7\%.) Statistical errors in $aM_{\rm
kin}$ are of order 1\%.

(3) {\it Tuning errors in the bare quark mass.} Since $a^{-1}$ is known
only to $\pm 4\%$, the $\Upsilon$'s kinetic mass can only be tuned with an
accuracy of $\pm 4\%$. However our determination of~$M_b$
(\eq{smeth}) minimizes sensitivity to such tuning, as is
evident from Table~\ref{second-method}. Thus tuning errors
have a negligible effect on~$M_b$.
 \begin{table}
 \begin{center}
  \begin{tabular}{ccccccc|ccc}
  $\beta$ & $n_f$ & $aM_b^0$ & $aM_\kin(\Upsilon)$ &
  $b_m$ & $aq_m$ & $Z_m-1$ & $M_\kin(\Upsilon)$ & $M_b$ & $M_b^0$ \\ \hline
  6.0&0 & 1.71 & 3.94(3) & 0.48 & 0.49 & 0.20 &  9.5(4) & 4.92 & 4.11\\
      && 1.80 & 4.09(3) & 0.46 & 0.51 & 0.18 &  9.8(4) & 4.93 & 4.16 \\
      && 2.00 & 4.49(4) & 0.42 & 0.54 & 0.16 & 10.7(4) & 4.88 & 4.21 \\
      && 3.00 & 6.57(7) & 0.24 & 0.52 & 0.09 & 15.8(6) & 4.73 & 4.32 \\
  5.6 & 2& 1.80 & 4.17(2) & 0.46 & 0.51 & 0.23 & 10.0(4) & 5.02 & 4.08
\end{tabular}
 \end{center}
  \caption{Simulation and perturbative results used in the second method for
  determining~$M_b$. Values for $M_\kin(\Upsilon)$, $M_b$ and $M_b^0$
  (last column) are in
  GeV. Values for~$M_b$ are determined using~Eq.~(9). Values for
  $M_b^0$ are obtained using Eq.~(9) with $Z_m\!\to\!1$.
  Results are given for $n_f=0$ and 2 light-quark flavors, and for
  a range of bare quark masses; the
  $b$-quark has $aM_b^0 = 1.7(1)$. }
 \label{second-method}
 \end{table}

In this paper, we have presented two different determinations of the
$b$-quark mass, each consistent with a pole mass of
$M_b=5.0(2)$~GeV. The systematic errors are
quite different in each case, and have been probed by computing with a
variety of quark masses and quark lagrangians, with and without light-quark
vacuum polarization. The complete agreement between the two methods is a
strong indication of the validity of each, and more generally of the lattice
QCD techniques on which they rely.  In both cases the dominant uncertainties
are in the perturbative calculations. A two-loop evaluation of either the
quark energy~$E_0$ or the mass renormalization constant~$Z_m$ would probably
reduce the errors in~$M_b$ by  a factor of two or more.
We are currently examining the feasibility of such calculations.

Combining the standard result for the pole mass in terms of
the $\msbar$~mass~\cite{msbar-mass},
\be
M_b = M_b^\msbar (M_b)\, \left\{ 1 + 0.424\,\alpha_V(0.22 M_b) +
0.164\,\alpha_V^2 +\cdots \right\},
\ee
with our result in terms of
the bare quark mass, $M_b = Z_m\,M_b^0$, we obtain a
relation between the bare quark mass on the lattice and the $\msbar$~mass:
\be
M_b^\msbar(M_b) = M_b^0\,\left\{ 1 + 0.06 \alpha_V(q_\msbar) + \cdots\right\},
\ee
where $q_\msbar\approx 0.8/a$ for our $M_b^0$'s.
Using this relation, our data implies that $M_b^\msbar(M_b) =
4.0(1)$~GeV. $\order(\alpha_V^2)$ corrections should be smaller
than in \eq{smeth} for the pole mass. This is because the scale for
$\alpha_V$ is significantly larger here and therefore $\alpha_V$ is smaller.
The larger scale is sensible since the $\msbar$~mass is a
bare mass and so should be  less infrared sensitive than
the on-shell pole mass.

Our $b$-quark mass can be used to estimate the $c$-quark's pole mass. Because
of heavy-quark symmetry, the difference between the spin-averaged mass of
the~$B$ and $B^*$~mesons, $M_{B/B^*}$, and $M_b$ is the same as the difference
between the spin-averaged $D/D^*$ mass and $M_c$ up to corrections of
order~$\Lambda_{\rm QCD}^2/M_c$. Thus we expect
\be
M_c \approx M_{D/D^*} + \left(M_b - M_{B/B^*}\right) = 1.6(2)~{\rm GeV}.
\ee
A different estimate is possible if the binding energies of the~$\Upsilon$
and~$\psi$ are roughly equal. This is probably the case since radial and
orbital excitation energies are roughly equal in the two systems. Also
our simulation shows that the binding energy is almost mass
independent for masses larger than~$M_b$. Thus we expect
\be
M_c \approx (1/2)\,\left( M_{\psi} + (2M_b - M_{\Upsilon}) \right)
  = 1.8(2)~{\rm GeV}.
\ee
We are currently extending our lattice analysis to allow a direct
determination of $M_c$.

We thank Greg Anderson and Paul Mackenzie for discussions
concerning this analysis. The simulations described here were carried out at
the Ohio Supercomputer Center. This work was supported by grants from
the NSF, the SERC, and the DOE (DE-FG02-91ER40690, DE-FC05-85ER250000,
DE-FG05-92ER40742, DE-FG05-92ER40722).


\begin{thebibliography}{9}
\bibitem{kilcup} The gauge-field configurations used were generously
provided to us by Greg Kilcup and his collaborators.
\bibitem{unquenched} A full discussion of these results will be presented
elsewhere.
\end{thebibliography}
\end{document}